\documentclass[12pt]{article}

\usepackage{float}
\usepackage{amsfonts}
\usepackage{amsmath}
\usepackage{amsbsy}
\usepackage{graphicx}
\usepackage{amssymb}
\usepackage{amsthm}
\usepackage{bm}
\usepackage{epsfig}
\usepackage{latexsym}
\usepackage{pdflscape}
\usepackage{color}
\numberwithin{equation}{section}
\allowdisplaybreaks

\DeclareMathOperator{\tr}{tr}

\DeclareMathOperator{\Li}{Li}

\newcounter{aff}

\begin{document}

\begin{titlepage}
\begin{flushright}
{\footnotesize YITP-14-75}
\end{flushright}
\begin{center}
{\Large\bf ABJM Membrane Instanton\\[6pt]
from Pole Cancellation Mechanism
}

\bigskip\bigskip
{\large 
Sanefumi Moriyama\footnote[1]{\tt moriyama@math.nagoya-u.ac.jp}
\quad and \quad
Tomoki Nosaka\footnote[2]{\tt nosaka@yukawa.kyoto-u.ac.jp}
}\\
\bigskip
${}^{*}$\,
{\small\it Kobayashi Maskawa Institute
\& Graduate School of Mathematics, Nagoya University\\
Nagoya 464-8602, Japan}
\medskip\\
${}^{*\dagger}$\,
{\small\it Yukawa Institute for Theoretical Physics,
Kyoto University\\
Kyoto 606-8502, Japan}
\end{center}

\begin{abstract}
The coefficients of the membrane instantons in the ABJM theory are known to be quadratic polynomials of the chemical potential.
We show that, after deforming the ABJM theory into more general superconformal Chern-Simons theories labelled by $(q,p)$ where the original ABJM theory corresponds to $(q,p)=(1,1)$, we can decompose the membrane instanton into three types of non-perturbative effects with constant coefficients independent of the chemical potential.
We find that, although these constants contain poles at certain values of $q$ and $p$ including the ABJM case, all of the poles cancel among themselves and the finite quadratic polynomial coefficients are reproduced at these values.
This is similar to what happens between the membrane instantons and the worldsheet instantons in the ABJM theory.
\end{abstract}
\end{titlepage}

\section{Introduction}\label{intro}

Recently there is much progress in understanding the worldvolume theory of multiple M2-branes.
It was found in \cite{ABJM} that the worldvolume theory of $N$ coincident M2-branes on a geometry ${\mathbb C}^4/{\mathbb Z}_k$ is described by the ${\mathcal N}=6$ $U(N)_k\times U(N)_{-k}$ superconformal Chern-Simons theory where the subscripts $k$ and $-k$ denote the Chern-Simons levels associated to each $U(N)$ factor.
Also, many other superconformal Chern-Simons theories with less supersymmetries describe worldvolume theories of multiple M2-branes on less supersymmetric backgrounds \cite{GY,GW,HLLLP1,JT}.
After applying the localization theorem \cite{P,KWY}, the infinite-dimensional path integral in defining the partition function of these theories on $S^3$ is reduced to a finite-dimensional matrix integral.

One of the most remarkable results in the study of the partition function of the ABJM theory $Z_\text{ABJM}(N)$ on $S^3$ is the determination of the coefficients of the membrane instantons.
The membrane instanton was first introduced in \cite{BBS}.
In the context of the ABJM theory, the membrane instanton is the non-perturbative effect interpreted as an M2-brane wrapping a three-dimensional submanifold in $\mathbb{C}^4/\mathbb{Z}_k$ \cite{DMP2}.
Subsequently, the coefficients of the instanton effects were explicitly determined.
If we define the grand potential $J_\text{ABJM}(\mu)$ as
\begin{align}
e^{J_\text{ABJM}(\mu)}=\sum_{N=0}^\infty Z_\text{ABJM}(N)e^{\mu N},
\end{align}
by introducing the chemical potential $\mu$ dual to $N$, the membrane instantons are explicitly given by \cite{MP}
\begin{align}
J_\text{ABJM}^\text{MB}(\mu)
=\sum_{\ell=1}^\infty
(a_\ell(k)\mu^2+b_\ell(k)\mu+c_\ell(k))e^{-2\ell\mu},
\label{MBinst}
\end{align}
where $a_\ell(k)$, $b_\ell(k)$ and $c_\ell(k)$ are $\mu$-independent constants given in \cite{HMO3,HMMO}.
For example, the explicit form of the coefficients of the first membrane instanton is given with \cite{HMO2,CM}
\begin{align}
a_1(k)=-\frac{4\cos\frac{\pi k}{2}}{\pi^2 k},\quad
b_1(k)=\frac{2\cos^2\frac{\pi k}{2}}{\pi\sin\frac{\pi k}{2}},\quad
c_1(k)=\frac{\pi}{6}\biggl(1+\frac{k^2}{8}\biggr)a_1(k)-\frac{k^2}{2}\frac{\partial}{\partial k}\biggl(\frac{b_1(k)}{k}\biggr).
\label{abc1}
\end{align}

In the standard situations, an instanton coefficient is usually a constant.
In contrast to it, it is perplexing to find that the coefficients of \eqref{MBinst} are quadratic polynomials of the chemical potential.
The fact of the coefficients being polynomials may suggest that the membrane instanton contains some further structures to be clarified.

Some clues to this puzzle were already found in the developments so far.
The first one is the so-called pole cancellation mechanism \cite{HMO2} used to determine the expression of \eqref{MBinst}.
Let us first recapitulate it.
Following many interesting aspects of the ABJM matrix model \cite{MPtop,DMP1,HKPT,FHM,O}, it was discovered \cite{MP} that we can regard the partition function as that of a non-interacting ideal Fermi gas system with $N$ particles which are governed by a non-trivial one-particle Hamiltonian,
\begin{align}
e^{-\widehat H_\text{ABJM}}
=\frac{1}{2\cosh\frac{\widehat Q}{2}}\frac{1}{2\cosh\frac{\widehat P}{2}},
\end{align}
with the Planck constant in the canonical commutation relation $[\widehat Q,\widehat P]=i\hbar$ given by $\hbar=2\pi k$.
In terms of this Hamiltonian, the grand potential is given by
\begin{align}
J_\text{ABJM}(\mu)
=\sum_{\ell=1}^\infty\frac{(-1)^{\ell-1}e^{\ell\mu}}{\ell}
\tr e^{-\ell\widehat H_\text{ABJM}}.
\end{align}
This Fermi gas formalism is not only suitable for the systematic WKB $\hbar$ expansion \cite{MP,CM}, but also applicable to the study of the exact values of the partition function \cite{HMO1,PY} which lead directly to the numerical results of the grand potential \cite{HMO2,HMO3}.
Combining with the results from the 't Hooft genus expansion \cite{DMP1,FHM,KEK} and the dual description through the topological string theory on local ${\mathbb P}^1\times{\mathbb P}^1$ \cite{MPtop}, finally the whole large $\mu$ expansion of the grand potential including the non-perturbative terms were written down explicitly \cite{HMMO}.
It was found that the non-perturbative effects in the grand potential consist of two types of instantons and their bound states.
One is the worldsheet instanton
\begin{align}
J_\text{ABJM}^\text{WS}(\mu)
=\sum_{m=1}^\infty d_m(k)e^{-\frac{4m\mu}{k}},
\label{WSinst}
\end{align}
which can be described by the free energy of the topological string theory on local ${\mathbb P}^1\times{\mathbb P}^1$ \cite{MPtop,DMP1,HMO2}.
Here the exponential factor $e^{-\frac{4\mu}{k}}$ is interpreted as a fundamental string wrapping $\mathbb{CP}^1$ \cite{CSW,DMP1} in the IIA picture.
The other is the membrane instanton \eqref{MBinst} where the exponential factor $e^{-2\mu}$ is interpreted as a D2-brane wrapping $\mathbb{RP}^3$ \cite{DMP2}.
After the whole studies of the partition function of the ABJM theory, one of the main conclusions is that the membrane instanton is described by the free energy of the refined topological string theory in the Nekrasov-Shatashvili limit \cite{NS} on the same background \cite{HMMO}.\footnote{
Some further studies such as the spectral problem, the perturbation series and the special supersymmetry enhancements can be found in \cite{KM,HW,GMZ,CGM,WWH,GHM}.
}

In the determination of these non-perturbative effects, the so-called pole cancellation mechanism \cite{HMO2} played a crucial role.
It was found \cite{HMO2} that the coefficients of the worldsheet instanton \eqref{WSinst} contain poles at certain values of $k$.
Since the matrix model itself takes finite values, these poles must be cancelled by those from other non-perturbative contributions.
If we assume that the coefficients of the membrane instantons also have the poles thus required, we finally obtain the exact expressions of the coefficients of the membrane instantons, which are consistent with the WKB $\hbar$ expansion \cite{MP,CM} and reproduce the numerical results of \cite{HMO2,HMO3} after the pole cancellation.
Furthermore, if we adopt the free energy of the refined topological strings in the Nekrasov-Shatashvili limit for the membrane instantons, we can see \cite{HMMO} that all of the poles from the free energy of the topological strings describing the worldsheet instantons are cancelled.
In this sense, we can say that the whole membrane instantons are determined by the pole cancellation mechanism. 

The second clue is the appearance of two types of membrane instantons in the generalizations of the ABJM theory.
It is interesting to ask how general it is that the pole cancellation mechanism can determine the non-perturbative expansions.\footnote{
For a generalization to the case of two different ranks $U(N_1)_k\times U(N_2)_{-k}$ \cite{HLLLP2,ABJ}, see \cite{AHS,H,MM,HoOk,K}.
}
In our previous work \cite{MN} we proceeded to more general ${\mathcal N}=4$ superconformal Chern-Simons theories of the circular quiver type \cite{IK4} with the levels given by\footnote{
A special case of the ${\cal N}=4$ theories called orbifold ABJM theory \cite{BKKS,IK,TY} was studied in \cite{HM}.
Also, a similar analysis on a closely related model \cite{GM} in a slightly different language, which corresponds to the $\{s_a\}_{a=1}^M=\{(+1)^{N_f},(-1)\}$ case in our language, can be found in \cite{HaOk}, which appeared almost simultaneously as \cite{MN}.
}
\begin{align}
k_a=\frac{k}{2}(s_a-s_{a-1}),\quad s_a=\pm 1.
\label{N4level}
\end{align}
The Fermi gas formalism is also applicable to this class of theories and the Hamiltonian is given by
\begin{align}
e^{-\widehat H}
=\frac{1}{(2\cosh\frac{\widehat Q}{2})^{q_1}}
\frac{1}{(2\cosh\frac{\widehat P}{2})^{p_1}}
\frac{1}{(2\cosh\frac{\widehat Q}{2})^{q_2}}
\frac{1}{(2\cosh\frac{\widehat P}{2})^{p_2}}\cdots
\frac{1}{(2\cosh\frac{\widehat Q}{2})^{q_m}}
\frac{1}{(2\cosh\frac{\widehat P}{2})^{p_m}},
\end{align}
for $\{s_a\}_{a=1}^M=\{(+1)^{q_1},(-1)^{p_1},(+1)^{q_2},(-1)^{p_2},\cdots,(+1)^{q_m},(-1)^{p_m}\}$.
Here this expression denotes a sequence consisting of $q_1$ elements of $+1$, $p_1$ terms of $-1$ and so on in this ordering.
For the perturbative part and the membrane instanton part, we fully utilized the WKB $\hbar$ expansion for this Fermi gas system
\begin{align}
J^\text{pert+MB}(\mu)=\sum_{n=1}^\infty\hbar^{n-1}J_n(\mu).
\label{WKB}
\end{align}
After analyzing the first few terms in the $\hbar$ expansion, we detected two types of non-perturbative effects where the exponential factors are given by $e^{-\frac{2\mu}{q}}$ and $e^{-\frac{2\mu}{p}}$ with 
\begin{align}
q=\sum_{a=1}^mq_a,\quad p=\sum_{a=1}^mp_a.
\label{sa}
\end{align}
As the exponents are independent of $k$, we expected that they can be interpreted as generalizations of the membrane instantons.

Now let us come back to the original puzzle, the quadratic polynomial coefficients in the membrane instantons.
From these two clues, if we introduce two deformation parameters $(q,p)$, it is natural to expect that the ABJM membrane instanton \eqref{MBinst} splits into two or more fundamental non-perturbative effects with constant coefficients containing poles at certain values of $(q,p)$, and that the polynomial coefficients in \eqref{MBinst} appear after cancelling these poles.
In fact, in this paper we shall see that this is the case.

The setup in this paper is as follows.
We shall study the minimal generalization $\{s_a\}_{a=1}^M=\{(+1)^q,(-1)^p\}$ with general values of $(q,p)$, which reduces to the ABJM case for $(q,p)=(1,1)$.
We consider only the WKB expansion of the membrane instanton around $k=0$.
The grand potential in this case was found to be \cite{MN}
\begin{align}
J_0(\mu)&=\sum_{\ell=1}^\infty\frac{(-1)^{\ell-1}e^{\ell\mu}}{\ell}\int\frac{dQdP}{2\pi}\frac{1}{(2\cosh\frac{Q}{2})^{q\ell}}\frac{1}{(2\cosh\frac{P}{2})^{p\ell}}\nonumber \\
&=\sum_{\ell=1}^\infty
\frac{(-1)^{\ell-1}e^{\ell\mu}}{2\pi\ell}
\frac{\Gamma(\frac{q\ell}{2})^2}{\Gamma(q\ell)}
\frac{\Gamma(\frac{p\ell}{2})^2}{\Gamma(p\ell)},
\label{J0}
\end{align}
without much change from the ABJM case \cite{MP}.

Let us summarize our main results.
Although the original definition \eqref{J0} is given in the small $e^\mu$ expansion, if we generalize $(q,p)$ to irrational numbers, we can rewrite it into the large $e^\mu$ expansion, where aside from the perturbative term
\begin{align}
J_0^\text{pert}(\mu)=\frac{4}{3\pi qp}\mu^3+\frac{\pi(4-q^2-p^2)}{3qp}\mu+\frac{2(q^3+p^3)}{\pi qp}\zeta(3),
\label{J0pert}
\end{align}
we have the following three kinds of non-perturbative terms,
\begin{align}
J_0^{(q)}(\mu)&=\sum_{m=1}^\infty\begin{pmatrix}2m\\m\end{pmatrix}
\frac{1}{m\sin\frac{2\pi m}{q}}
\frac{\Gamma(-\frac{pm}{q})^2}{\Gamma(-\frac{2pm}{q})}
e^{-\frac{2m\mu}{q}},\nonumber\\
J_0^{(p)}(\mu)&=\sum_{n=1}^\infty\begin{pmatrix}2n\\n\end{pmatrix}
\frac{1}{n\sin\frac{2\pi n}{p}}
\frac{\Gamma(-\frac{qn}{p})^2}{\Gamma(-\frac{2qn}{p})}
e^{-\frac{2n\mu}{p}},
\label{J0qp}
\end{align}
and
\begin{align}
J_0^{(2)}(\mu)=\sum_{l=1}^\infty\frac{(-1)^{l-1}}{2\pi l}
\frac{\Gamma(-\frac{ql}{2})^2}{\Gamma(-ql)}
\frac{\Gamma(-\frac{pl}{2})^2}{\Gamma(-pl)}e^{-l\mu}.
\label{J02}
\end{align}
Note that the coefficients of the non-perturbative effects are not quadratic polynomials any more but constants independent of the chemical potential.
After taking the deformation parameters $(q,p)$ back to $(1,1)$ for the ABJM theory, we encounter various poles.
After cancelling all the poles, we come back to the original quadratic polynomials of the ABJM theory.
This indicates that we have decomposed the original membrane instanton of the ABJM theory into more fundamental ones.
We hope that our analysis is helpful not only in studying the ${\cal N}=4$ theories, but also in understanding the original membrane instanton in the ABJM theory itself.

The remaining part of this paper is organized as follows.
In section \ref{3sectors}, we shall rewrite the small $e^\mu$ expansion of the grand potential \eqref{J0} into the large $e^\mu$ expansion, where we find three types of non-perturbative effects \eqref{J0qp} and \eqref{J02}.
Although the coefficients contain poles at various values of $(q,p)$, all of the poles cancel among themselves to reproduce the quadratic polynomials, as we shall see in section \ref{cancellation}.
In section \ref{J2J4}, we apply our large $\mu$ expansion to the subsequent orders in the WKB $\hbar$ expansion.
We conclude in section \ref{conclusion} with discussions.

\section{From small $e^\mu$ to large $e^\mu$}\label{3sectors}

The grand potential in the classical limit $\hbar\to 0$, $J_0(\mu)$, is obtained as a power series in $e^{\mu}$ \eqref{J0}, which is appropriate at $\mu\rightarrow-\infty$.
In this section, we shall rewrite this series into a large $\mu$ expansion to derive the perturbative part \eqref{J0pert} and the non-perturbative corrections \eqref{J0qp}, \eqref{J02} in $J_0(\mu)$.
Below we generalize $q$ and $p$ to be irrational numbers, to avoid any divergences which possibly appear.

We first introduce numerical constants $\gamma_m$ defined by
\begin{align}
\frac{\Gamma(x)^2}{\Gamma(2x)}
=\frac{2}{2^{2x}}\sum_{m=0}^\infty\frac{\gamma_m}{m+x},\quad
\gamma_m=\frac{1}{2^{2m}}\begin{pmatrix}2m\\m\end{pmatrix}.
\label{g2/g}
\end{align}
Using these constants, the power series expansion of $J_0(\mu)$ with respect to $e^\mu$ \eqref{J0} is rewritten into
\begin{align}
J_0(\mu)=-\frac{8}{\pi qp}
\sum_{\ell=1}^\infty(-e^{\mu^\prime})^\ell
\sum_{m=0}^\infty\sum_{n=0}^\infty
\frac{\gamma_m\gamma_n}
{\ell\bigl(\ell+\frac{2m}{q}\bigr)\bigl(\ell+\frac{2n}{p}\bigr)}.
\label{J0rew}
\end{align}
Here we have introduced
\begin{align}
\mu^\prime=\mu-(q+p)\log 2
\end{align}
for abbreviation.
Using the partial fraction decomposition, we find that the coefficient in the summand is written as
\begin{align}
&\sum_{m=0}^\infty\sum_{n=0}^\infty
\frac{\gamma_m\gamma_n}{\ell(\ell+\frac{2m}{q})(\ell+\frac{2n}{p})}
=\sum_{m=1}^\infty\sum_{n=0}^\infty
\frac{q^2\gamma_m\gamma_n}{4m^2(1-\frac{nq}{mp})}
\frac{1}{\ell+\frac{2m}{q}}
+\sum_{m=0}^\infty\sum_{n=1}^\infty
\frac{p^2\gamma_m\gamma_n}{4n^2(1-\frac{mp}{nq})}
\frac{1}{\ell+\frac{2n}{p}}
\nonumber\\
&\quad+\sum_{m=1}^\infty\sum_{n=1}^\infty
\frac{qp\gamma_m\gamma_n}{4mn}\frac{1}{\ell}
+\sum_{m=1}^\infty \gamma_m\biggl(\frac{q}{2m}\frac{1}{\ell^2}
-\frac{q^2}{4m^2}\frac{1}{\ell}\biggr)
+\sum_{n=1}^\infty \gamma_n\biggl(\frac{p}{2n}\frac{1}{\ell^2}
-\frac{p^2}{4n^2}\frac{1}{\ell}\biggr)
+\frac{1}{\ell^3},
\label{fracdec}
\end{align}
where we have used $\gamma_0=1$.
Now let us perform the summation over $\ell$ in \eqref{J0rew}.
To obtain the large $\mu$ expansion, we use the formulae
\begin{align}
\sum_{\ell=1}^\infty\frac{(-e^\mu)^\ell}{\ell+\alpha}
=-\frac{1}{\alpha}+\frac{\pi}{\sin\pi\alpha}e^{-\alpha\mu}
-\sum_{\ell=1}^\infty\frac{(-e^{\mu})^{-\ell}}{-\ell+\alpha},
\label{poleformula}
\end{align}
and
\begin{align}
\Li_1(-e^\mu)&=-\mu+\Li_1(-e^{-\mu}),\quad
\Li_2(-e^\mu)=-\frac{\mu^2}{2}-\frac{\pi^2}{6}-\Li_2(-e^{-\mu}),
\nonumber\\
\Li_3(-e^\mu)&=-\frac{\mu^3}{6}-\frac{\pi^2\mu}{6}+\Li_3(-e^{-\mu}),
\label{polylogformula}
\end{align}
for the polylogarithm function
\begin{align}
\Li_s(z)&=\sum_{\ell=1}^\infty\frac{z^\ell}{\ell^s}.
\label{polylogdef}
\end{align}
Here all of these formulae \eqref{poleformula} and \eqref{polylogformula} can be derived from\footnote{
It is interesting to note that the same formula with $\mu$ purely imaginary was used in the light-cone string field theory \cite{CG,GS} to prove the unitarity \cite{KMT,KiMo} of the overlapping matrices.
}
\begin{align}
\sum_{\ell=-\infty}^\infty\frac{(-e^\mu)^\ell}{\ell+\alpha}
&=\frac{\pi}{\sin\pi\alpha}e^{-\alpha\mu}.
\end{align}
With the help of these formulae, we divide $J_0(\mu)$ into four parts: the perturbative terms and the non-perturbative terms of $e^{-\frac{2\mu}{q}}$, $e^{-\frac{2\mu}{p}}$, $e^{-\mu}$.

First let us consider the non-perturbative terms of $e^{-\frac{2\mu}{q}}$ and $e^{-\frac{2\mu}{p}}$, which are collected as
\begin{align}
J_0^{(q)}(\mu)&=2\sum_{m=1}^\infty\sum_{n=0}^\infty
\frac{\gamma_m\gamma_n}{m\bigl(n-\frac{mp}{q}\bigr)\sin\frac{2\pi m}{q}}
e^{-\frac{2m\mu^\prime}{q}},\nonumber \\
J_0^{(p)}(\mu)&=2\sum_{n=1}^\infty\sum_{m=0}^\infty
\frac{\gamma_m\gamma_n}{n\bigl(m-\frac{nq}{p}\bigr)\sin\frac{2\pi n}{p}}
e^{-\frac{2n\mu^\prime}{p}}.
\end{align}
In these expressions, we can perform the summation over $n$ in $J_0^{(q)}(\mu)$ (or over $m$ in $J_0^{(p)}(\mu)$) just by the definition \eqref{g2/g}, and we finally obtain \eqref{J0qp}.

Next we consider the non-perturbative terms of $e^{-\mu}$, which are
\begin{align}
J_0^{(2)}(\mu)&=\frac{8}{\pi qp}\sum_{\ell=1}^\infty
(-e^{\mu^\prime})^{-\ell}
\biggl[\sum_{m=1}^\infty\sum_{n=0}^\infty
\frac{q^2\gamma_m\gamma_n}
{4m^2(1-\frac{nq}{mp})}\frac{1}{-\ell+\frac{2m}{q}}
+\sum_{m=0}^\infty\sum_{n=1}^\infty
\frac{p^2\gamma_m\gamma_n}{4n^2(1-\frac{mp}{nq})}\frac{1}{-\ell+\frac{2n}{p}}
\nonumber\\
&\hspace{-1cm}-\sum_{m=1}^\infty\sum_{n=1}^\infty
\frac{qp\gamma_m\gamma_n}{4mn}\frac{1}{\ell}
+\sum_{m=1}^\infty \gamma_m
\biggl(\frac{q}{2m}\frac{1}{\ell^2}+\frac{q^2}{4m^2}\frac{1}{\ell}\biggr)
+\sum_{n=1}^\infty \gamma_n
\biggl(\frac{p}{2n}\frac{1}{\ell^2}+\frac{p^2}{4n^2}\frac{1}{\ell}\biggr)
-\frac{1}{\ell^3}
\biggr].
\label{J02reverse}
\end{align}
This expression of $J_0^{(2)}(\mu)$ seems lengthy.
However, we can compute it without much effort.
First we notice that the expression \eqref{J02reverse} is obtained by using \eqref{poleformula} and \eqref{polylogformula}.
The formula \eqref{poleformula} converts the $e^{\mu}$ terms into the $e^{-\mu}$ terms just by replacing $\ell$ with $-\ell$ and simultaneously changing the overall signs.
This is the case also for \eqref{polylogformula} if we substitute the power series expression of the polylogarithm function \eqref{polylogdef}.
This observation means that $J_0^{(2)}(\mu)$ can be computed by using the formula \eqref{fracdec} inversely, with the same flips of signs
\begin{align}
J_0^{(2)}(\mu)=\frac{8}{\pi qp}\sum_{\ell=1}^\infty(-e^{\mu^\prime})^{-\ell}\sum_{m=0}^\infty\sum_{n=0}^\infty\frac{\gamma_m\gamma_n}{-\ell\bigl(-\ell+\frac{2m}{q}\bigr)\bigl(-\ell+\frac{2n}{p}\bigr)}.
\end{align}
Summing over $m$ and $n$ by \eqref{g2/g}, one ends up with \eqref{J02}.

Finally we consider the perturbative terms,
\begin{align}
&J_0^\text{pert}(\mu)=
-\frac{q}{\pi p}\sum_{m=1}^\infty\sum_{n=0}^\infty
\frac{\gamma_m\gamma_n}{m^2\bigl(\frac{n}{p}-\frac{m}{q}\bigr)}
-\frac{p}{\pi q}\sum_{n=1}^\infty\sum_{m=0}^\infty
\frac{\gamma_m\gamma_n}{n^2\bigl(\frac{m}{q}-\frac{n}{p}\bigr)}
\nonumber \\
&\quad+\biggl(\frac{2\bar{\gamma}_1^2}{\pi}
-\frac{2\bar{\gamma}_2}{\pi}\biggl(\frac{q}{p}+\frac{p}{q}\biggr)
\biggr)\mu^\prime
+\frac{4\bar{\gamma}_1}{\pi}\biggl(\frac{1}{p}+\frac{1}{q}\biggr)
\biggl(\frac{\mu^{\prime 2}}{2}+\frac{\pi^2}{6}\biggr)
+\frac{8}{\pi qp}\biggl(\frac{\mu^{\prime 3}}{6}
+\frac{\pi^2\mu^\prime}{6}\biggr).
\end{align}
Here we have introduced other numerical constants
\begin{align}
\bar{\gamma}_s=\sum_{m=1}^\infty\frac{\gamma_m}{m^s},
\end{align}
whose explicit values are
\begin{align}
\bar{\gamma}_1=2\log 2,\quad
\bar{\gamma}_2=\frac{\pi^2}{6}-2(\log 2)^2,\quad
\bar{\gamma}_3=-\frac{\pi^2\log 2}{3}+\frac{4(\log 2)^3}{3}+2\zeta(3).
\label{ctilde}
\end{align}
To calculate this expression, note that the first two terms sum up to
\begin{align}
\frac{\bar{\gamma}_3}{\pi}\biggl(\frac{q^2}{p}+\frac{p^2}{q}\biggr)
-\frac{\bar{\gamma}_1\bar{\gamma}_2(q+p)}{\pi}.
\end{align}
Plugging this in, with the explicit values of $\bar{\gamma}_s$ \eqref{ctilde}, we obtain the result \eqref{J0pert}.

\section{Pole cancellation mechanism}\label{cancellation}

In the previous section, we have seen the large $\mu$ expansion of the classical limit of the grand potential $J_0(\mu)$.
We have found that the large $\mu$ expansion contains three types of non-perturbative contributions $e^{-\frac{2\mu}{q}}$, $e^{-\frac{2\mu}{p}}$ and $e^{-\mu}$ respectively in $J_0^{(q)}(\mu)$, $J_0^{(p)}(\mu)$ and $J_0^{(2)}(\mu)$ with coefficients being constant independent of the chemical potential $\mu$.
There we have extrapolated $(q,p)$ into general irrational numbers to obtain the results \eqref{J0qp} and \eqref{J02}.
These resulting expressions indicate that, in the case of integral $(q,p)$, which is our original interest, the coefficient of each sector contains divergent contributions.

In this section, we shall see that these divergences completely cancel among themselves.
The cancellation is indeed consistent, since the grand potential $J_0(\mu)$ \eqref{J0} itself is well-defined for arbitrary positive $(q,p)$.
Remarkably, the coefficients in the non-perturbative effects remaining after these pole cancellations are generally polynomials in $\mu$.

In the following, we first rewrite the results into a symmetric expression which is suitable for seeing how the pole cancellation occurs.
Then, restricting ourselves to the cases where all the three sectors contribute to the cancellation (which is the only possibility for the ABJM theory), we explicitly write down the general form of the remaining coefficients.
We obtain quadratic polynomials in these cases, which exactly coincide with the previously obtained ones for the ABJM theory \cite{MP,CM} and the ${\cal N}=4$ theories \cite{MN}.
Finally we see an implication of the form of these quadratic polynomials.

\subsection{A symmetric expression}

To simplify the discussion of the pole cancellation, let us first rewrite the three sectors of non-perturbative contributions, $J_0^{(q)}(\mu)$, $J_0^{(p)}(\mu)$ and $J_0^{(2)}(\mu)$, into an expression symmetric under the exchange of $q$, $p$ and $2$.
We find that they can be expressed as\footnote{
In the discovery of this expression, we are partially stimulated by some previous works.
In \cite{N}, the $n$-ple sine function is decomposed into $n$ sectors symmetric under the exchange of the $n$ parameters, each of which takes the form of the series expansion.
In \cite{LV}, the partition function on $S^5$ is expressed similarly.
Also in a note by Kazumi Okuyama, he was trying to formulate the cancellation mechanism between the membrane instantons and the worldsheet instantons in the analogy of these works.
}
\begin{align}
J_0^{(z_i)}(\mu)=\sum_{\ell_i=1}^\infty
\frac{F(\frac{\ell_i}{z_i};\mu)}{\ell_i}
\prod_{j=1(\neq i)}^3\cot\frac{\pi z_j\ell_i}{z_i},
\label{J0instsym}
\end{align}
where we have introduced $z_i=(q,p,2)$, $\ell_i=(m,n,l)$ and
\begin{align}
F(r;\mu)=-\frac{2\pi}{\cos 2\pi r}
\frac{\Gamma(2qr+1)}{\Gamma(qr+1)^2}
\frac{\Gamma(2pr+1)}{\Gamma(pr+1)^2}e^{-2r\mu}.
\label{Fqp}
\end{align}
Indeed it is not difficult to find that each sector in \eqref{J0instsym} reduces to \eqref{J0qp} and \eqref{J02} after the substitution $(z_1,z_2,z_3)=(q,p,2)$.
In the derivation, we need to flip the signs in the arguments of the Gamma functions using
\begin{align}
\Gamma(x)\Gamma(1-x)=\frac{\pi}{\sin\pi x}.
\label{gamsine}
\end{align}

In the expression \eqref{J0instsym} all of the Gamma functions in the coefficients are free from divergence, while the cotangent factors imply that each sector contains the non-perturbative effects with divergent coefficients.
Explicitly speaking, the divergence appears
at $m\in\frac{q}{\gcd(q,p)}\mathbb{N}\cup\frac{q}{\gcd(q,2)}\mathbb{N}$
in $J_0^{(q)}(\mu)$,
at $n\in\frac{p}{\gcd(p,2)}\mathbb{N}\cup\frac{p}{\gcd(p,q)}\mathbb{N}$
in $J_0^{(p)}(\mu)$
and at $l\in\frac{2}{\gcd(2,q)}\mathbb{N}\cup\frac{2}{\gcd(2,p)}\mathbb{N}$
in $J_0^{(2)}(\mu)$.
However, as $F(r;\mu)$ from different sectors share the same instanton exponent at these points, we expect that the divergences are cancelled among those terms with the same exponent.
By replacing $(q,p)$ with $(q(1+\varepsilon_1),p(1+\varepsilon_2))$ to regularize the divergences and taking the limit $\varepsilon_1,\varepsilon_2\rightarrow 0$ after summing all the contributions, we find that our expectation is indeed correct.
In the next subsection, as an example, we demonstrate this in detail for the cancellation among the three sectors and determine the finite coefficients remaining after the cancellation.

\subsection{Cancellation among three sectors}

When the instanton numbers of the three sectors $(m,n,l)$ satisfy
\begin{align}
\frac{m}{q}=\frac{n}{p}=\frac{l}{2}\,(=:r),
\label{3pcancel}
\end{align}
where $r\in{\mathbb N}/\gcd(q,p,2)$, all the three sectors contribute to the non-perturbative effect of $e^{-2r\mu}$.

Let us see how the pole cancellation works.
For this purpose, we substitute $z_i(1+\varepsilon_i)$ for $z_i$ and send $\varepsilon_i\to 0$.
Note that we do not have to introduce $\varepsilon_3$ to shift $z_3=2$ in discussing the cancellation.
The cancellation becomes clearer, however, by introducing $\varepsilon_3$ and treating three $z_i$ on the equal footing.
For simplicity, we introduce the notation
\begin{align}
F_\varepsilon(r;\mu)=F(r;\mu)|_{q\rightarrow q^\prime,p\rightarrow p^\prime}
\end{align}
with $q^\prime=q(1+\varepsilon_1)$, $p^\prime=p(1+\varepsilon_2)$ and leave $q^\prime$ and $p^\prime$ in $F_\varepsilon(r;\mu)$ untouched while expanding other factors around $\varepsilon_i\rightarrow 0$.
Then we find that the term in $J_0^{(z_i)}(\mu)$ contributing to $e^{-2r\mu}$ is
\begin{align}
&F_\varepsilon\biggl(\frac{\ell_i}{z_i(1+\varepsilon_i)};\mu\biggr)
\frac{1}{\ell_i}
\cot\frac{\pi z_j\ell_i}{z_i}\frac{1+\varepsilon_j}{1+\varepsilon_i}
\cot\frac{\pi z_k\ell_i}{z_i}\frac{1+\varepsilon_k}{1+\varepsilon_i}
\nonumber\\
&=\biggl(F_\varepsilon(r;\mu)
-\frac{\varepsilon_ir}{1+\varepsilon_i}\partial F_\varepsilon(r;\mu)
+\frac{\varepsilon_i^2r^2}{2(1+\varepsilon_i)^2}
\partial^2 F_\varepsilon(r;\mu)
+{\cal O}(\varepsilon^3)\biggr)\nonumber \\
&\quad\times\frac{1}{z_ir}
\biggl(\frac{1+\varepsilon_i}{\pi z_jr\varepsilon_{ji}}
-\frac{1}{3}\frac{\pi z_jr\varepsilon_{ji}}{1+\varepsilon_i}
+{\mathcal O}(\varepsilon^3)\biggr)
\biggl(\frac{1+\varepsilon_i}{\pi z_kr\varepsilon_{ki}}
-\frac{1}{3}\frac{\pi z_kr\varepsilon_{ki}}{1+\varepsilon_i}
+{\mathcal O}(\varepsilon^3)\biggr),
\end{align}
where $j$, $k$ denote the two indices\footnote{
The readers should not confuse the index $k$ appearing only in this subsection with the Chern-Simons level $k=\hbar/2\pi$.
}
other than $i$ and we have introduced the shorthand notation $\varepsilon_{ji}=\varepsilon_j-\varepsilon_i$.
Collecting the terms which formally scale in the non-positive powers in $\varepsilon$, we find
\begin{align}
\frac{F_\varepsilon(r;\mu)}{\pi^2r^3z_1z_2z_3}
\frac{(1+\varepsilon_i)^2}{\varepsilon_{ji}\varepsilon_{ki}}
-\frac{\partial F_\varepsilon(r;\mu)}{\pi^2r^2z_1z_2z_3}
\frac{\varepsilon_i(1+\varepsilon_i)}{\varepsilon_{ji}\varepsilon_{ki}}
+\frac{\partial^2F_\varepsilon(r;\mu)}{2\pi^2rz_1z_2z_3}
\frac{\varepsilon_i^2}{\varepsilon_{ji}\varepsilon_{ki}}
-\frac{F_\varepsilon(r;\mu)}{3rz_1z_2z_3}
\biggl(\frac{z_k^2\varepsilon_{ki}}{\varepsilon_{ji}}
+\frac{z_j^2\varepsilon_{ji}}{\varepsilon_{ki}}\biggr).
\end{align}
With the help of the identities
\begin{align}
\sum_{i=1}^3\frac{1}{\varepsilon_{ji}\varepsilon_{ki}}=0,
\quad
\sum_{i=1}^3\frac{\varepsilon_i}{\varepsilon_{ji}\varepsilon_{ki}}=0,
\quad
\sum_{i=1}^3\frac{\varepsilon_i^2}{\varepsilon_{ji}\varepsilon_{ki}}=1,
\end{align}
we can show that the terms in formally negative power of $\varepsilon$ vanish after summed over all the three sectors.\footnote{
Note that the terms of formally positive power in $\varepsilon$ simplify into a homogeneous polynomial of that degree.
For example, the terms proportional to $(\epsilon_{ji}\epsilon_{ki})^{-1}$ sum up to the Schur polynomial $(n>2)$
\begin{align}
\sum_{i=1}^3
\frac{\varepsilon_i^n}{\varepsilon_{ji}\varepsilon_{ki}}=\chi_{(n-2)}(\varepsilon_1,\varepsilon_2,\varepsilon_3).
\end{align}
This fact guarantees that these contributions vanish in the limit of $\varepsilon_i\rightarrow 0$, regardless of the direction of the limit.
}
Because of it, we can safely change $F_\varepsilon(r;\mu)$ back to $F(r;\mu)$.
Finally, the finite part is given by
\begin{align}
\frac{F(r;\mu)-r\partial_r F(r;\mu)
+\frac{1}{2}r^2\partial^2_rF(r;\mu)}{\pi^2z_1z_2z_3r^3}
-\frac{(z_1^2+z_2^2+z_3^2)F(r;\mu)}{3z_1z_2z_3r}.
\label{finite}
\end{align}

Calculating $F(r;\mu)$ and its derivatives, with the explicit form of $F(r;\mu)$ in \eqref{Fqp}, we finally obtain the following contribution of the non-perturbative effects $e^{-2r\mu}$
\begin{align}
&\frac{F(r;\mu)}{2\pi^2qpr^3}
\biggl[2r^2\mu^2+\bigl(2r-4r^2{H}_{1}(r)\bigr)\mu\nonumber\\
&+1-2r{H}_{1}(r)
+r^2\bigl(2{H}_1(r)^2-{H}_2(r)\bigr)
+\frac{\pi^2r^2(4-q^2-p^2)}{6}\biggr].
\label{explicit2dpol}
\end{align}
Here ${H}_s(r)$ is defined with the harmonic numbers
\begin{align}
h_s(m)=\sum_{\ell=1}^m\frac{1}{\ell^s},
\end{align}
as
\begin{align}
{H}_s(r)=q^s\bigl(2^{s-1}h_s(2qr)-h_s(qr)\bigr)
+p^s\bigl(2^{s-1}h_s(2pr)-h_s(pr)\bigr).
\label{Hqp}
\end{align}
These ${H}_s(r)$ result from the derivatives of the Gamma functions in $F(r;\mu)$, using the formula
\begin{align}
\psi^{(0)}(m)=-\gamma+h_1(m-1),\quad \psi^{(1)}(m)=\frac{\pi^2}{6}-h_2(m-1),
\end{align}
where $\gamma$ is the Euler-Mascheroni constant and the polygamma functions are define as
\begin{align}
\psi^{(s-1)}(x)=\biggl(\frac{d}{dx}\biggr)^s\log\Gamma(x).
\end{align}
As we have expected in section \ref{intro}, quadratic polynomial coefficients have appeared in \eqref{explicit2dpol} as a result of the pole cancellation.

This explicit form indeed reproduces the previous results in the ${\cal N}=4$ theories of $\{s_a\}=\{(+1)^q,(-1)^p\}$ \cite{MN} which were obtained by expressing the grand potential $J_0(\mu)$ with the generalized hypergeometric function $_{q+p+2}F_{q+p+1}(e^{2\mu^\prime})$ where $q,p$ were the numbers of the parameters and should be integers throughout the analysis.
Especially, with $q=p=1$, the membrane instanton coefficients in the limit $k\rightarrow 0$ in the ABJM theory \cite{MP,CM} are reproduced.

At the poles where only two of the three sectors contribute, on the other hand, we obtain linear polynomials in $\mu$ as the remaining finite parts.
These are again consistent with the results obtained in \cite{MN}.

\subsection{Effective chemical potential}\label{secmueff}

As a byproduct, in this subsection we shall discuss an implication of the expressions \eqref{explicit2dpol} for general ${\cal N}=4$ theories.
Let us express the results for the WKB expansion \eqref{WKB} schematically as
\begin{align}
J^\text{pert+MB}(\mu)
=\frac{C}{3}\mu^3+B\mu+A+J_a(\mu)\mu^2+J_b(\mu)\mu+J_c(\mu).
\end{align}
Here $A$, $B$ and $C$ are perturbative coefficients.
The explicit form of $C$ \cite{GHP} and $B$ \cite{MN} is
\begin{align}
C=\frac{4}{\pi\hbar qp},\quad
B=\frac{1}{\pi}\biggl(\frac{\hbar qp}{48}
+\pi^2\frac{4-q^2-p^2}{3\hbar qp}\biggr),
\end{align}
while the explicit form of $A$ is not used below.
On the other hand, the non-perturbative contributions $J_a(\mu)$, $J_b(\mu)$ and $J_c(\mu)$ are given by $(r\in\mathbb{N}/\gcd(q,p,2))$
\begin{align}
&J_a(\mu)
=\frac{1}{\pi\hbar}\sum_{r} a_r e^{-2r\mu}+{\cal O}(\hbar),
\quad
J_b(\mu)
=\frac{1}{\pi\hbar}\sum_{r} b_r e^{-2r\mu}+\cdots+{\cal O}(\hbar),\nonumber \\
&J_c(\mu)=\frac{1}{\pi\hbar}\sum_{r}
(c_r+\pi^2c_r^\prime)e^{-2r\mu}+\cdots+{\cal O}(\hbar),
\label{rationality}
\end{align}
where all of the coefficients $a_r$, $b_r$, $c_r$ and $c_r^\prime$ are rational numbers whose explicit forms are given in \eqref{explicit2dpol}.
Note that there are also non-perturbative contributions with different exponents in $J_b(\mu)$ and $J_c(\mu)$, though they do not affect the argument in this subsection.
In the case of the ABJM theory, it was found \cite{HMO3} that the large $\mu$ expansion simplifies extensively if we redefine the chemical potential $\mu$ into
\begin{align}
\mu_\text{eff}=\mu+\frac{J_a(\mu)}{C}.
\label{mueff}
\end{align}
Indeed, the worldsheet instanton part takes care of all the bound states of the worldsheet instanton and the membrane instanton; the quadratic part of the instanton coefficients is completely absorbed into the perturbative part; the $c_\ell^\prime$ terms are also absorbed and the $c_\ell$ terms are the derivatives of $b_\ell$.
In this subsection we shall find that in the redefinition $\mu_\text{eff}(\mu)$ in a general $(q,p)$ model, one of the simplifications, the cancellation of the $c_\ell^\prime$ terms, still takes place.

In fact, in terms of $\mu_\text{eff}$, it is not difficult to find that the linear part and the constant part are shifted as
\begin{align}
{\widetilde J}_b(\mu_\text{eff})=J_b(\mu)-\frac{J_a(\mu)^2}{C},\quad
{\widetilde J}_c(\mu_\text{eff})=J_c(\mu)-\frac{J_a(\mu)J_b(\mu)}{C}-\frac{BJ_a(\mu)}{C}+\frac{2J_a(\mu)^3}{3C^2},
\end{align}
Now we find that not only the coefficients in ${\widetilde J}_b(\mu_\text{eff})$ but also those in ${\widetilde J}_c(\mu_\text{eff})$ are rational numbers except the overall factor $1/\pi$.
Indeed the terms in $\pi{\widetilde J}_c(\mu_\text{eff})$ proportional to $\pi^2$, coming only from $J_c(\mu)$ and $-BJ_a(\mu)/C$, completely cancel as
\begin{align}
c_{r}^\prime-\frac{B}{C}\cdot a_{r}=\frac{r^2(4-q^2-p^2)}{6}-\frac{\frac{4-q^2-p^2}{3\hbar qp}}{\frac{4}{\hbar qp}}\cdot 2r^2=0.
\label{irrcancel}
\end{align}
Remarkably, this cancellation of irrationality is also true for the higher $\hbar$ corrections, as we explain at the end of the next section.

In the ABJM theory, the introduction of the effective chemical potential $\mu_\text{eff}$ was important as we have explained above.
This non-trivial rationality in the coefficients of non-perturbative contributions might imply that the effective chemical potential also plays an important role in the ${\cal N}=4$ theories.

\section{Higher order corrections}\label{J2J4}

So far we have considered the grand potential $J_0(\mu)$ in the leading order of the classical limit $\hbar\to 0$.
In this section, we shall consider the higher order correction in $\hbar$ to the grand potential.
We shall see that our results for $J_0(\mu)$ obtained in the previous sections are straightforwardly generalized to these corrections.

In \cite{MN}, we found that, introducing a generalization of the power series \eqref{J0},
\begin{align}
{\cal F}(\alpha,\beta;\mu)
&=\sum_{\ell=1}^\infty\frac{(-1)^{\ell-1}e^{\ell\mu}}{\ell}
\int\frac{dQdP}{2\pi}
\frac{1}{\bigl(2\cosh\frac{Q}{2}\bigr)^{q\ell+\alpha}}
\frac{1}{\bigl(2\cosh\frac{P}{2}\bigr)^{p\ell+\beta}}\nonumber\\
&=\sum_{\ell=1}^\infty\frac{(-1)^{\ell-1}e^{\ell\mu}}{2\pi\ell}
\frac{\Gamma(\frac{q\ell+\alpha}{2})^2}{\Gamma(q\ell+\alpha)}
\frac{\Gamma(\frac{p\ell+\beta}{2})^2}{\Gamma(p\ell+\beta)},
\label{calFgen}
\end{align}
with $\alpha$ and $\beta$ being non-negative even integers, then, as well as the leading order $J_0(\mu)={\cal F}(0,0;\mu)$, the $\hbar$ corrections $J_2(\mu)$ and $J_4(\mu)$ to the grand potential \eqref{WKB} are also expressed in terms of ${\cal F}(\alpha,\beta;\mu)$ as
\begin{align}
J_2(\mu)&=\frac{qp}{24}(1-\partial_\mu^2){\mathcal F}(2,2;\mu),
\nonumber\\
J_4(\mu)&=\frac{(qp)^2}{5760}\Bigl[-(1-\partial_\mu^2)(9-\partial_\mu^2)f_{41}+(1-\partial_\mu^2)(4-\partial_\mu^2)f_{42}\Bigr],
\label{Jhigher}
\end{align}
with
\begin{align}
f_{41}={\mathcal F}(4,4;\mu)
+\frac{1}{2}{\mathcal F}(2,4;\mu)
+\frac{1}{2}{\mathcal F}(4,2;\mu)
+\frac{1}{4}{\mathcal F}(2,2;\mu),\quad
f_{42}={\mathcal F}(2,2;\mu).
\end{align}

If we continue $q$ and $p$ to irrational numbers, we can obtain the large $\mu$ expansion of the function ${\cal F}(\alpha,\beta;\mu)$ by the same method used in section \ref{3sectors}.
In the current case, instead of \eqref{g2/g}, the expansion of the ratio of the Gamma functions reads
\begin{align}
\frac{\Gamma(x+\frac{\alpha}{2})^2}{\Gamma(2x+\alpha)}
=\frac{2}{2^{2x+\alpha}}
\sum_{m=\frac{\alpha}{2}}^\infty
\frac{\gamma_{m-\frac{\alpha}{2}}}{m+x},
\end{align}
and, instead of \eqref{fracdec}, for $\alpha,\beta\ge 2$ the partial fraction decomposition is simply
\begin{align}
\frac{1}{\ell(\ell+\frac{2m}{q})(\ell+\frac{2n}{p})}
=\frac{qp}{4mn}\frac{1}{\ell}
-\frac{q}{2m(\frac{2n}{p}-\frac{2m}{q})}\frac{1}{\ell+\frac{2m}{q}}
-\frac{p}{2n(\frac{2m}{q}-\frac{2n}{p})}\frac{1}{\ell+\frac{2n}{p}}.
\end{align}

We finally obtain the large $\mu$ expansion of ${\cal F}(\alpha,\beta;\mu)$ which consists of, other than the perturbative parts,
\begin{align}
{\cal F}^\text{pert}(\alpha,\beta;\mu)
=\frac{1}{2\pi}
\frac{\Gamma(\frac{\alpha}{2})^2}{\Gamma(\alpha)}
\frac{\Gamma(\frac{\beta}{2})^2}{\Gamma(\beta)}
\Bigl[\mu-q\bigl(\psi^{(0)}(\alpha)-\psi^{(0)}(\alpha/2)\bigr)
-p\bigl(\psi^{(0)}(\beta)-\psi^{(0)}(\beta/2)\bigr)\Bigr],
\end{align}
the three non-perturbative parts
\begin{align}
{\mathcal F}^{(z_i)}(\alpha,\beta;\mu)
&=\sum_{\ell_i=\lambda_i}^\infty
\frac{F_{(\alpha,\beta)}(\frac{\ell_i}{z_i};\mu)}{\ell_i}
\prod_{j=1(\ne i)}^3
\cot\frac{\pi z_j\ell_i}{z_i}.
\label{Fgeninstsym}
\end{align}
Here we have defined $(\lambda_1,\lambda_2,\lambda_3)=(\frac{\alpha}{2},\frac{\beta}{2},1)$ and
\begin{align}
F_{(\alpha,\beta)}(r;\mu)=-\frac{2\pi}{\cos 2\pi r}
\frac{\Gamma(2qr-\alpha+1)}{\Gamma(qr-\frac{\alpha}{2}+1)^2}
\frac{\Gamma(2pr-\beta+1)}{\Gamma(pr-\frac{\beta}{2}+1)^2}
e^{-2r\mu},
\label{Fgen}
\end{align}
In the derivation, we have used \eqref{gamsine} to change the arguments of the Gamma functions as previously. 

Roughly speaking, the pole cancellation works in the same way as in the case of $\alpha=\beta=0$ discussed in section \ref{cancellation}: terms from different sectors share the same instanton exponent at the point where the cotangent factors diverge.
The main difference is that the pole cancellation among the three sectors happens at $(m,n,l)=(qr,pr,2r)$ with $r\in{\mathbb N}/\gcd(q,p,2)$, only when the instanton number is large enough to satisfy $m\ge\frac{\alpha}{2}$ and $n\ge\frac{\beta}{2}$.
Finally, the finite part remaining after the cancellation is given by
\begin{align}
\frac{F_{(\alpha,\beta)}(r;\mu)
-r\partial_r F_{(\alpha,\beta)}(r;\mu)
+\frac{1}{2}r^2\partial_r^2F_{(\alpha,\beta)}(r;\mu)}{\pi^2z_1z_2z_3r^3}
-\frac{(z_1^2+z_2^2+z_3^2)F_{(\alpha,\beta)}(r;\mu)}{3z_1z_2z_3r},
\end{align}
or explicitly, as a quadratic polynomial in $\mu$ 
\begin{align}
&\frac{F_{(\alpha,\beta)}(r;\mu)}{2\pi^2qpr^3}
\biggl[2r^2\mu^2+\bigl(2r-4r^2{H}_{1(\alpha,\beta)}(r)\bigr)\mu
\nonumber \\
&+1-2r{H}_{1(\alpha,\beta)}(r)
+r^2\bigl(2{H}_{1(\alpha,\beta)}(r)^2-{H}_{2(\alpha,\beta)}(r)\bigr)
+\frac{\pi^2r^2(4-q^2-p^2)}{6}\biggr].
\label{2dpolgen}
\end{align}
Here we define the generalization of $H_s(r)$ in \eqref{Hqp}, $H_{s(\alpha,\beta)}(r)$ as
\begin{align}
H_{s(\alpha,\beta)}(r)
=q^s\Bigl(2^{s-1}h_s(2qr-\alpha)
-h_s\Bigl(qr-\frac{\alpha}{2}\Bigr)\Bigr)
+p^s\Bigl(2^{s-1}h_s(2pr-\beta)
-h_s\Bigl(pr-\frac{\beta}{2}\Bigr)\Bigr),
\label{Hgen}
\end{align}
which again comes from the derivatives of the Gamma functions in $F_{(\alpha,\beta)}(r;\mu)$.

For the small instanton number, we have to be careful, since the corresponding contribution from ${\cal F}^{(q)}(\alpha,\beta;\mu)$ or from ${\cal F}^{(p)}(\alpha,\beta;\mu)$ sometimes do not exist due to the lower bounds on the instanton number, $m\ge\frac{\alpha}{2}$ and $n\ge\frac{\beta}{2}$.
At the first sight it might seems that we have too many divergent cotangent factors to obtain the finite result.
In these cases, however, the ratio of the Gamma functions becomes zero, which reduces the power of divergences.
This can also be seen from the expression before the rewriting using \eqref{gamsine}.

In subsection \ref{secmueff}, we have discussed the simplification of the non-perturbative effects of $e^{-2r\mu}$ with $r\in\mathbb{N}/\gcd(q,p,2)$ associated to the redefinition of the chemical potential \eqref{mueff}.
In the discussion there, the following properties of the coefficient \eqref{explicit2dpol} are essential:
the rationality of $a_r,b_r,c_r,c^\prime_r$ in \eqref{rationality} and the $r$-independence of the ratio of $a_r$ and $c^\prime_r$ \eqref{irrcancel}.
As we have claimed in subsection \ref{secmueff}, the same simplification occurs also in the higher $\hbar$ corrections. 
Here we shall see it explicitly by showing these properties.
Since these properties are preserved under the differential operations in \eqref{Jhigher} which convert ${\cal F}(\alpha,\beta;\mu)$ to $J_n(\mu)$, we have only to care the coefficients of the non-perturbative effects in ${\cal F}(\alpha,\beta;\mu)$ themselves.
For the case of the large instanton number, $m\ge\frac{\alpha}{2}$ and $n\ge\frac{\beta}{2}$, the coefficients in ${\cal F}(\alpha,\beta;\mu)$ are given by \eqref{2dpolgen} and these properties can be explicitly checked as for $J_0(\mu)$ in subsection \ref{secmueff}.
For the case where one of these two conditions is not satisfied, the result \eqref{2dpolgen} is no longer valid.
However, we can see $a_r=c^\prime_r=0$.
First, since the divergence is at most ${\cal O}(\varepsilon^{-1})$, as argued in the paragraph below \eqref{Hgen}, the second derivative of $F_{(\alpha,\beta)}(r;\mu)$ does not appear and thus $a_r=0$.
Secondly, the relative $\pi^2$ factor would only appear in the second derivative of $F_{(\alpha,\beta)}(r;\mu)$ or in the cross terms of ${\cal O}(\varepsilon^{-1})$ and ${\cal O}(\varepsilon)$ between two cotangent factors.
Since both of these terms are absent in this case, $c_r^\prime$ is also zero.
Moreover, the explicit calculation shows the rationality of the other two, $b_r$ and $c_r$.
Therefore the required properties hold also in this case.

There is still another way to obtain the large $\mu$ expansion of the function ${\cal F}(\alpha,\beta;\mu)$.
From the power series definition \eqref{calFgen}, we find that the following differential relations are satisfied
\begin{align}
(q\partial_\mu+\alpha+1){\cal F}(\alpha+2,\beta;\mu)
&=\frac{1}{4}(q\partial_\mu+\alpha){\cal F}(\alpha,\beta;\mu),
\nonumber\\
(p\partial_\mu+\beta+1){\cal F}(\alpha,\beta+2;\mu)
&=\frac{1}{4}(p\partial_\mu+\beta){\cal F}(\alpha,\beta;\mu).
\label{recursion}
\end{align}
Decomposing these equations further into those for the terms with the same instanton exponents, we obtain the recursion relation between the coefficient in ${\cal F}(\alpha+2,\beta;\mu)$ (or in ${\cal F}(\alpha,\beta+2;\mu)$) and the corresponding one in ${\cal F}(\alpha,\beta;\mu)$.
Regarding the constant coefficient in the non-perturbative sectors of $J_0(\mu)={\cal F}(0,0;\mu)$ in \eqref{J0instsym} as the initial value for the recursion relation, we can reproduce the results for ${\cal F}(\alpha,\beta;\mu)$ in \eqref{Fgeninstsym}.
In passing let us note that we can also use the relation \eqref{recursion} to obtain the perturbative part or the polynomial coefficients of the non-perturbative effects remaining after the pole cancellation.

To summarize our analysis for the higher order corrections, we find that the total grand potential $J^\text{pert+MB}(\mu)$ in the WKB expansion obtained so far are given by
\begin{align}
J^\text{pert+MB}(\mu)=\biggl(\frac{1}{\hbar}{\cal D}_0+\hbar{\cal D}_2+\hbar^3{\cal D}_4\biggr)J_0(\mu)+{\cal O}(\hbar^5),
\end{align}
with
\begin{align}
{\cal D}_0&=1,\quad
{\cal D}_2=\frac{q^2p^2(1-\partial_\mu^2)\partial_\mu^2}
{384(1+q\partial_\mu)(1+p\partial_\mu)},\nonumber\\
{\cal D}_4&=\frac{q^3p^3(1-\partial_\mu^2)\partial_\mu^2}
{92160(1+q\partial_\mu)(1+p\partial_\mu)}
\biggl(-\frac{(9-\partial_\mu^2)(8+3q\partial_\mu)(8+3p\partial_\mu)}
{16(3+q\partial_\mu)(3+p\partial_\mu)}+4-\partial_\mu^2\biggr).
\end{align}
Here we have used the recursion relation \eqref{recursion} to relate ${\cal F}(\alpha,\beta;\mu)$ to ${\cal F}(0,0;\mu)$.
For the non-perturbative effects with constant coefficients, each $\partial_\mu$ is replaced with $-2m/q$, $-2n/p$ or $-l$.
We hope that this expression is helpful in determining the coefficients of the membrane instantons at finite $k$.

\section{Conclusion and discussion}\label{conclusion}

In this paper we have obtained a new understanding of the coefficients of the membrane instantons in the ABJM theory.
First, the ABJM matrix model is generalized to include two parameters $q$ and $p$.
Due to these deformation parameters, the membrane instantons are subdivided into three instanton sectors, whose coefficients are $\mu$-independent constants while are singular in the undeformed limit $q,p\rightarrow 1$.
The quadratic polynomial coefficients of the membrane instantons in the ABJM theory emerge as a result of the pole cancellation among these sectors.

Though we do not have concrete field theoretic realization of the instanton effects, we expect that this decomposition will provide us a better physical interpretation of the membrane instantons in the ABJM theory.
In view of the standard interpretation of the instanton coefficient as the volume of the instanton moduli space, we are tempted to give a similar interpretation to our results.
From this viewpoint the divergence might denote the non-compactness of the instanton moduli space, while the cancellation implies the non-perturbative compactification of the moduli space.

It is reasonable for a skeptical reader to ask whether our decomposition of the membrane instantons in the ABJM theory into three in the deformed theory (with irrational $q,p$) really helps us to understand them.
As already happened in the ABJM theory, however, it is only after we deformed the integral Chern-Simons level $k$ into an irrational number that we were able to split the non-perturbative effects into the worldsheet instantons and the membrane instantons, and describe them in terms of the refined topological strings.
We expect that the irrationality of $q,p$ will play a similar role.

Of course, to solve the ${\cal N}=4$ circular quiver superconformal Chern-Simons theories is itself an interesting future work.
From a technical viewpoint we have also made progress in this direction.
If we denote as $(q,p)_k$ the ${\mathcal N}=4$ theories with the circular quivers where the Chern-Simons levels \eqref{N4level} are given by $\{s_a\}_{a=1}^M=\{(+1)^q,(-1)^p\}$, we can summarize recent progress by saying that the first few instanton coefficients of the $(N_f,1)_1$ model and the $(2,1)_k$ model were studied via the pole cancellation mechanism in \cite{HaOk} and \cite{MN} respectively.
In contrast to these one-parameter deformations, let us stress that our current work is the first one which succeeds in studying the model with the two-parameter deformation, $(q,p)_0$.

Though we have considered only the $\{s_a\}_{a=1}^M=\{(+1)^q,(-1)^p\}$ cases where the different signs of $s_a$ are completely separated in the circular quiver, the explicit expansion of $J_0(\mu)$ is valid also for general ${\mathcal N}=4$ theories with \eqref{sa}, since the ordering of operators is irrelevant in the strictly classical limit.
Furthermore, for the higher order corrections, our argument on the pole cancellation among the membrane instantons can be straightforwardly extended.
This is because the grand potential obtained in the WKB expansion can be generally expressed by using ${\cal F}(\alpha,\beta;\mu)$ in \eqref{calFgen}, as observed in \cite{MN}.

The ${\cal N}=4$ theory with $s_a$ satisfying \eqref{sa} for some $q$ and $p$ is dual to the eleven dimensional supergravity on $\text{AdS}_4\times S^7/\Gamma$, where $\Gamma$ is generated by three non-independent operations $\mathbb{Z}_k$, $\mathbb{Z}_{q}$ and $\mathbb{Z}_{p}$, with the discrete torsion \cite{IK,IY}.
We hope to understand the three kinds of non-perturbative effects as membranes wrapping submanifolds in $S^7/\Gamma$, as in the case of the ABJM theory \cite{DMP2}.
Especially, our explicit calculation indicates that, although there are bound states of the worldsheet instantons and the membrane instantons, there are no bound states among the three types of the membrane instantons without the worldsheet instantons.
Also, if we claim that there are three types of the membrane instantons $e^{-\frac{2\mu}{q}}$, $e^{-\frac{2\mu}{p}}$, $e^{-\mu}$, when $q$, $p$ are odd integers, we expect that $e^{-\mu}$ with odd instanton numbers can be distinguished from the other two.
However, the constant coefficients of these effects always vanish.
This is why we could not detect it in our previous work \cite{MN}.
We hope to explain these observations from the supergravity analysis in the future study.

It would also be interesting to apply our idea to understand the results obtained in \cite{HM,HaOk}.
The grand potential obtained in \cite{HM} for $r=4$, $k=2$ and the one obtained in \cite{HaOk} with $k=3,6$ contain polynomials of degree higher than 2 in instanton coefficients.
As we have commented in the introduction, the matrix model considered in \cite{HaOk} can also be realized in the setup of the ${\cal N}=4$ circular quiver, where the coefficients of the membrane instantons are at most quadratic.
Therefore, these results imply that, in the more general theories than the ABJM theory, the non-perturbative effects have more abundant fine structures to be clarified.

\section*{Acknowledgements}

We are grateful to Yasuyuki Hatsuda, Shinji Hirano, Hiroaki Kanno, Fuminori Nakata, Shin Nayatani, Kazumi Okuyama, Sara Pasquetti, Hidehiko Shimada, Fumihiko Sugino, Sotaro Sugishita, Alessandro Torrielli, Alexander Varchenko for valuable discussions.
The work of S.M.\ is partly supported by JSPS Grant-in-Aid for Scientific Research (C) \#26400245, while the work of T.N.\ is partly supported by the JSPS Research Fellowships for Young Scientists.

\end{document}